\documentclass{article}
\usepackage[T1]{fontenc}
\usepackage{graphicx}
\usepackage{epsf,rotate}
\usepackage{epsfig}
\usepackage{rotating}
\makeatletter

\newcommand{\LyX}{L\kern-.1667em\lower.25em\hbox{Y}\kern-.125emX\spacefactor1000}

\usepackage[T1]{fontenc}

\makeatletter

\usepackage[T1]{fontenc}
\usepackage{geometry}


\makeatother

\begin{document}

\title{Study of Damage Propagation at the 
Interface Localization-Delocalization Transition of the Confined Ising Model.}
\author{M. Leticia Rubio Puzzo and Ezequiel V. Albano}
\date{Instituto de Investigaciones Fisicoqu\'{\i}micas Te\'{o}ricas y 
Aplicadas (INIFTA), UNLP, CONICET. Casilla de Correo 16, 
Sucursal 4, (1900) La Plata, Argentina.}
\maketitle
%\date{today}

\begin{abstract}
The propagation of damage in a confined magnetic Ising film, 
with short range competing magnetic fields ($h$) acting 
at opposite walls, is studied by means of Monte Carlo 
simulations. Due to the presence of the fields, the film 
undergoes a wetting transition at a well defined critical 
temperature $T_w(h)$. In fact, the competing 
fields causes the occurrence of an interface between magnetic 
domains of different orientation.
For $T < T_w(h)$ ($T > T_w(h)$)  such interface is bounded
(unbounded) to the walls, while right at $T_w(h)$ the 
interface is essentially located at the center of the film.  

It is found that the spatio-temporal spreading of the damage  
becomes considerably enhanced by the presence of the interface, 
which act as a ''catalyst'' of the damage causing an enhancement of the 
total damaged area. The critical points for damage spreading are 
evaluated by extrapolation to the thermodynamic limit using a
finite-size scaling approach. Furthermore, the wetting transition effectively
shifts the location of the damage spreading critical points, as compared
with the well known critical temperature of the order-disorder 
transition characteristic of the Ising model. Such a critical points
are found to be placed within the non-wet phase. 

\end{abstract}

%\vskip 1.0 true cm
	
%Keywords: Damage propagation in magnetic materials, 
%interfaces in magnetic systems, Monte Carlo numerical simulations.

%\vskip 1.0 true cm

PACS numbers: 75.70.-i, 75.30.Kz, 75.10.Hk, 05.10.Ln

\pagebreak

\section*{I. Introduction}

The study of the thermodynamic properties of confined systems 
has attracted considerable attention over the last decades
\cite{nak1,nak2,eva,evapi,parr1,swi,grad,land,parr2,bind1,kur,macio1,macio2,mamu,Kar,car,carlon,kur1,md,aign,ch,slen,tsay,albapa,scft,aarao}. 
The critical behavior
of confined fluids is rather different from the bulk
criticality due to the subtle interplay between finite-size
and surface effects. The interaction of a saturated gas
in contact with a wall or a substrate may result in the
occurrence of very interesting wetting and capillary condensation phenomena,
where a macroscopically thick liquid layer condenses
at the wall, while the bulk fluid may remain in the gas 
phase \cite{degene,diet,for,parry,sull}. The wetting of solid surfaces
by a fluid is a phenomena of primary importance in many fields
of practical technological applications (lubrication, efficiency 
of detergents, oil recovery in porous material, stability of 
paint coatings, interaction of macromolecules with interfaces, etc.
\cite{degene,Iglo,Ebn,Leb}). 

Wetting transitions are also observed when a magnetic
material is confined between parallel walls where competing surface 
magnetic fields act. For example, when an Ising \cite{ising,Wu} 
film is confined between 
two competing walls a distance $L$ apart from each other, 
so that the surface magnetic fields ($H$) are of the same 
magnitude but opposite direction, it is found that
the competing fields cause the emergence of an interface that
undergoes a localization-delocalization transition. 
This transition shows up at an $L-$dependent temperature 
$T_{w}(L,H)$ that is the precursor of the true 
wetting transition temperature $T_{w}(H)$ of the infinite 
system \cite{eva,bind1,swi,abra1}. 
It should also be remarked that, although the 
discussion is presented here in terms of
a magnetic language, the relevant physical concepts can 
rather straightforwardly be extended 
to other systems such as fluids, polymers, and binary mixtures.  

On the other hand, the study and understanding of the 
propagation of perturbations in magnetic material is also
a subject of increasing interest. For this purpose,
the damage spreading method is a powerful and useful
technique that has been applied, for example, to Ising 
systems \cite{derrS,stan,grass1,montani,wang,vojta,lima1,lima2,alba4,alba55} 
as well as to spin glasses \cite{derrW,argo}, for a review see e.g.
\cite{herr}. 

In order to apply the damage spreading method one has to start 
from an equilibrium configuration of the magnetic material at 
temperature $T$. Such a configuration is  generically called the
reference or unperturbed configuration $S^{A}(T)$.
Subsequently, a perturbed configuration $S^{B}(T)$ is obtained
just introducing a small perturbation into $S^{A}(T)$. This procedure 
can be achieved, e.g. by flipping a small number of
spins of the unperturbed configuration. The time evolution of 
the perturbation is of primary interest because, either,
it can become healed after some time (eventually it may remain
finite) or it can propagated over the whole system. The
latter scenario may be undesired when a thermally stable magnetic
material needs to be achieved. A simple measure of the perturbation is
the ``Hamming distance'' or damage between the unperturbed
and the perturbed configurations \cite{derrS,stan}. 
Then the time evolution of the perturbation can be followed
evaluating the total damage $D(t)$ defined as
the fraction of spins with different orientations, that is
\begin{equation}
D(t)=\frac{1}{2N}\sum ^{N}_{l}\left| S^{A}_{l}(t,T)-S^{B}_{l}(t,T)]\right|,
\label{eq:dam}
\end{equation}
where the summation runs over the total number of spins $N$ and 
the index $l$, ($1\leq l\leq N$) is the label that identifies 
the spins of the configurations. 

In our previous study of damage spreading using Ising magnets confined 
in two-dimensional geometries \cite{alba4} and in absence of external 
magnetic fields, it has been shown that the presence of interfaces 
between magnetic domains, in the direction {\bf perpendicular} to the 
propagation of the damage, causes the spatio-temporal enhancement 
of the spreading. Due to this interesting feature and in order to 
contribute to the understanding of the role played by the interfaces, 
the aim of this work is to investigate the propagation of the damage 
at the interface generated by the localization-delocalization transition 
of the Ising magnet confined in presence of competing magnetic fields. 
In this case, and in contrast to the previous 
work \cite{alba4}, the propagation of the damage occurs in the
direction {\bf parallel} to the interface. In order to carry out the 
study,  Monte Carlo simulations of the Ising magnet in confined, 
thin film, geometries are performed. Imposing
open boundary conditions and surface magnetic fields along the surfaces
of the films, the propagation of the damage in both directions, parallel and
perpendicular, to the domain interfaces can be studied.
Furthermore, extrapolations of the results to the thermodynamic
limit allow us to determine the critical points of damage spreading
and locate them in the wetting phase diagram, as evaluated
exactly by Abraham \cite{abra1}.   

The manuscript is organized as follows: in Section II the numerical 
procedure for the simulation of damage spreading in confined 
geometries is described. Section III is devoted to a brief discussion 
of the equilibrium configurations characteristic of confined geometries 
with competing applied magnetic fields. The results are presented 
and discussed the results in Section IV,  while the conclusions 
are stated in Section V.

\section*{II. The confined Ising ferromagnet with competing fields and the 
Monte Carlo simulation method.}

The Ising model with competing surface
fields in a confined geometry of size $L\times M$ ($ L\ll M $) can be 
described by the following Hamiltonian $H$:
\begin{equation}
H=-J\sum ^{M,L}_{<ij,mn>}\sigma _{ij}\sigma _{mn}-h_{1}\sum ^{M}_{i=1}\sigma _{i1}-h_{L}\sum ^{M}_{i=1}\sigma _{iL} 
\label{eq:ham}
\end{equation}
where $\sigma _{ij}$ are the Ising spin variables, corresponding to the site
of coordinates $(i,j)$, that may assume two different values, 
namely $\sigma _{ij}=\pm 1$, $J>0$ is the coupling constant of the 
ferromagnet and the first summation of Equation (\ref{eq:ham}) runs over all 
the nearest-neighbor pairs of spins such as $1\leq i\leq M$ 
and $1\leq j\leq L$. The second (third) summation corresponds to 
the interaction of the spins placed at the surface layer $j=1$ ($j=L$) 
of the film where a short range surface magnetic field $h_1$ ($h_L$) acts. 
Open boundary conditions are assumed 
along the $M$-direction of the film where the fields act. 

In this manuscript, only the case
of competing surface fields such as $h_{1} = - h_{L}$, in the absence 
of any bulk magnetic field, is considered. So,  
hereafter $h = |h_{1}|= |h_{L}|$ will be used generically to
specify the short range surface magnetic field that is measured 
in units of the coupling constant $J$.  
Under these conditions and for a suitable range of fields and 
temperatures, one may observe the onset of an interface between
magnetic domains of opposite direction running along the film,
as will be discussed in detail below. 

The evolution of the Ising film is simulated using the Glauber 
dynamics, so a randomly selected spin is flipped with probability 
$p(flip)$ given by:
\begin{equation}
p(flip)=\frac{\exp (-\beta \cdot \bigtriangleup H)}{1+\exp (-\beta \cdot \bigtriangleup H)}
\label{eq:rates}
\end{equation}
where $\bigtriangleup H$ is the difference between the energy of the 
would-be new configuration and the old configuration, 
and $\beta =1/k_{B}T$ is the usual Boltzmann factor.

The time is measured in Monte Carlo time steps (mcs), such as during 
one mcs all $L\times M$ spins of the sample are flipped once, in 
the average.

The Ising magnet in two dimensions and in 
absence of any external magnetic field, undergoes a second-order 
order-disorder transition when the temperature is raised from a 
relatively low initial value. The critical point is the so called 
Onsager critical temperature $k_{B}T_{C}/J = 2.269...$ \cite{Wu}.

In order to evaluate the damage according to equation (\ref{eq:dam}), 
first the reference configuration $S^{A}$ 
is generated, using the simulation method described above.
This can be achieved starting from a random 
configuration and applying  the Glauber dynamics during $10^4$ mcs. 
Subsequently, a replica of such configuration is 
created and the spins of the central column ($i=\frac{M}{2}$) are 
flipped as follows: if the magnetization of the 
whole sample is greater than zero the up spins are flipped down, 
otherwise the down spins of the column are flipped up. Using this 
procedure the perturbed configuration $S^{B}$ of equation (\ref{eq:dam}) 
is generated and it is assured that the initial damage (at $t=0$) 
is always $D(t=0)\leq \frac{1}{M}$. This kind of perturbation 
reproduces the effect of a large magnetic field applied at the 
middle of the sample and pointing away to the opposite direction 
than the actual magnetization of the whole sample.

The time evolution of the damage D(t), evaluated
using equation (\ref{eq:dam}), is then followed applying the 
Glauber dynamics to both configurations. At this stage it is 
essential to use the same random numbers in both systems in 
order to perform the updates \cite{herr}. 

Applying the Glauber dynamics to the Ising ferromagnet,
it is known that exists a critical temperature for 
damage spreading ($T_{D}$), such as for 
$T>T_{D}$ the damage spreads out over the whole sample while 
for $T<T_D$ the damage becomes healed after 
some finite time \cite{herr}. Extensive numerical simulations due
to Grassberger \cite{grass2,grass3} have shown that, in two dimensions 
and in absence on external magnetic fields, $T_{D}\cong 0.992 T_{C}$.

\section*{III. Discussion of Equilibrium Configurations in 
Confined Geometries with Competing Fields.}

Since the aim of this manuscript is to investigate the role played 
by the interface between magnetic domains in the propagation of damage, 
it is worth discussing the different equilibrium configurations 
characteristic of confined geometries under different surface field 
conditions, close to the wetting phase transition.

The situation originated by the presence of competing surface 
magnetic fields $h_{1}=-h_{L}$ (see equation (\ref{eq:ham})) 
can be described in terms of a wetting transition that takes place 
at a certain -field dependent- critical wetting
temperature $T_{w}(h)$. In fact, for $T<T_w$ a small number of rows
parallel to one of the surface of the film have an overall 
magnetization pointing to the same direction than the adjacent 
surface field. However, the bulk of the film has the opposite 
magnetization (i.e. pointing in the direction of the
another competing field). Also the symmetric situation is equivalent
to the previous one due to the spin-reversal field-reversal symmetry.
These configurations, where the interface between domains is tightly bounded 
on the surface, are characteristic of the non-wet phase of the 
system that occurs at low enough temperatures and fields (see figure 1).
Increasing the temperature, such interface moves farther 
and farther away from the surface towards the bulk of the film. 
Just at $T_{w}$ the interface is located in the middle of the 
film and the system reaches the wet phase for the first time. For 
$T > T_{w}$ the interface is not longer localized and the system
is within the wet phase (see figure 1). 

It should be noticed that the above discussed wetting transition 
takes place in the thermodynamic limit only and the phase diagram
has been evaluated exactly by Abraham \cite{abra1}, yielding   
\begin{equation}
\exp(2J\beta) \cdot [\cosh(2J \beta) - \cosh(2h_{c}\beta)] = \sinh(2J \beta) ,
\label{eq:abrah}
\end{equation}
\noindent where $h_{c}(T)$ is the critical surface field  
(the inverse function of the wetting temperature $T_{w}(h)$).

As discussed above, a well defined wetting transition takes place 
in the thermodynamic limit only. Nevertheless, as shown in figure(2) 
a precursor of this wetting transition can also be observed in 
confined geometries for finite values of $L$. This precursor 
is most correctly described in terms of a localization-delocalization 
transition of the interface, which takes place at $L-$dependent
critical temperatures $(T_w(L,h))$. In fact, for $T<T_w(L,h)$ (figure 2(a)) 
one observes coexistence of two phases, each of them having 
opposite magnetization. Furthermore, the interface between domains 
is located close to one of the surfaces of the film 
(localized interface; non-wet regime). For $T > T_{w}(L,h)$ (figure 2(c)),
the wall between domains moves along the $L$-direction and 
the system enters to the wet regime (delocalized interface). 
The mean position of the interface remains close to the 
center of the film just at $T_{w}(L,h)$ (figure 2(b)). 

According to the finite-size scaling theory \cite{eva,swi} 
$T_{w}(L,h)$ shifts towards $T_{w}(h)$ according to:       
\begin{equation}
T_{w}(\infty,h) - T_{w}(L,h) \sim constant \times L^{-1} ,
\label{eq:Tfinsize}
\end{equation}
\noindent when the system size $L$ tends to the thermodynamic limit.
For further discussions on the localization-delocalization 
transition of the Ising system in confined geometries 
see e.g. \cite{eva,bind1}.

Finally, it is worth discussing the equilibrium configurations
obtained in absence of magnetic fields and close below to $T_{C}$, as 
shown in figure 2(d). In this case the confined system forms a 
nonuniform state given by a succession of up-spins and down-spins 
domains, which are essentially ordered 
at length scales such as the standard correlation length 
($\xi$) remains smaller than the characteristic size of
the system ($\xi < L$), but the domain walls are randomly placed.
The operation of surface magnetic fields favors the parallel
spins, and thus suppress this domain formation. These facts 
can be recognized from a visual inspection of the domain patterns
shown in figure 2.  

\section*{IV. Results and discussion.}

Monte Carlo simulations have been performed using $L \times M$-lattices, 
for the choices $L=12$, $24$, $48$ and $M=50 \times L$. 

Preliminary runs indicate that the damage spreading transition can be 
observed, in principle, for any range of temperatures and fields close 
to the critical wetting line as defined by equation (\ref{eq:abrah}). 
However, when the temperature is very low the system becomes almost frozen
making difficult the acquisition of data with reliable statistics.
Also, for $T > T_{C}$ the system is disordered 
preventing the occurrence of  the wetting transition. 
Taking into account these constraints,
an extensive study of damage spreading has been performed only 
close to two particular points of the whole wet 
non-wet phase diagram, namely 
for $\left| h_{1}\right| =\left| h_{L}\right| =0.5;  
T_{w}(h=0.5)\cong 0.863 T_{C}$, and for 
$T_{w} = 0.90 T_{C}; \left| h_{1}\right| =
\left| h_{L}\right| \cong 0.4291$. 
Additional studies have also been carried out close to
$T_{w} = 0.75 T_{C}; \left| h_{1}\right| =
\left| h_{L}\right| \cong 0.6652$ and 
$T_{w} = 0.95 T_{C}; \left| h_{1}\right| =
\left| h_{L}\right| \cong 0.3053$.

On view of the obtained results, it is 
expected that the 
main conclusions obtained from the study will hold close
to the whole critical wetting curve.

In order to characterize the global dynamics of the propagation 
of the perturbation, the whole damage given by the Hamming distance $D(t)$ 
(see equation (\ref{eq:dam})) and the survival probability $P(t)$, 
that is the probability that at time $t$ the damage is still propagating, 
have been evaluated. Also, to study the spatio-temporal evolution 
of the perturbation, the damage profile in the perpendicular 
(parallel) $P_{x}(i,t)$ ($P_{y}(i,t)$) direction is defined as follows:
\begin{equation}
P_{x}(i,t)=\frac{1}{2L}\sum^{L}_{j=1}\left| S^{A}_{i,j}(t,T) -
S^{B}_{i,j}(t,T)]\right|,
\label{eq:perfilx} 
\end{equation}
and
\begin{equation}
%\qquad 
P_{y}(j,t)=\frac{1}{2M}\sum ^{M}_{i=1}\left| S^{A}_{i,j}(t,T) - 
S^{B}_{i,j}(t,T)]\right|,
\label{eq:perfily}
\end{equation}
\noindent where $S^{A}_{i,j}(t,T)$ and $S^{B}_{i,j}(t,T)$ are the 
unperturbed and perturbed equilibrium configurations, respectively. 
This definition of the perpendicular (parallel) damage profile 
represents the average damaged sites of the $i-th$ - $i = 1,..,M$ column 
( $j-th$ - $j = 1,..,L$ row) of the lattice.
In order to characterize the profiles it is also useful to determine 
their width ($\langle \omega_{x}(t)\rangle$ and 
$\langle \omega_{y}(t)\rangle$), as well as their amplitude 
($\langle h_{x}(t)\rangle$ and $\langle h_{y}(t) \rangle$), 
respectively. 

The propagation of the damage is studied using two approaches:
i) keeping $h = 0.5 $ constant and varying $T$, and
ii) keeping $T = 0.90$ constant and varying $h$.

\subsection*{IV.1 The case $h = 0.5$, $T$ variable.}

Figure 3 shows log-log plots of the damage $D(t)$ as function of time. 
It is interesting to note that the localization-delocalization transition 
is roughly revealed by the behavior of the damage. In fact, for $T < T_w(L)$, 
(e.g. for $T = 0.70T_C$, $0.65T_C$ and $0.60T_C$ in figure 3(a) ) it is 
observed that the damage is healed after a relatively short time. 
However for $T > T_w(L)$ (e.g. for $T = 0.80T_C$ and $0.85T_C$ 
in figure 3(a)) the spreading of damage is actually observed. 

It should be noticed that using finite lattices, the onset
of damage spreading is observed at $L$-dependent ``critical''
temperatures ($T_{D}(L)$). As expected, finite-size effects cause
the actual transition to become rounded and shifted, as e.g.
can be observed comparing figures 3(a) and 3(b), which have been
obtained using lattices of different size.
The straight lines observed in figures 3(a) and 3(b) for $T \geq T_{D}(L)$ 
suggest a power-law behavior such as
\begin{equation}
D(t)\propto t^{\eta}
\label{eq:pwlaw}
\end{equation}
\noindent where \( \eta \) is an exponent. 
Disregarding the early time behavior of $D(t)$ (say up to $t \approx 30$ mcs), 
the best fit of the data shown in figure 3(a), using equation (\ref{eq:pwlaw}),
gives $\eta \cong 0.90\pm 0.02$. Simulations performed using 
bigger lattices ($L=24, 48$, see e.g. figure 3(b)) also 
give $\eta \cong 0.91\pm 0.01$, where
in all cases the error bars merely reflect the statistical error.

Notice that damage spreading  is also observed for a temperature 
slightly below $T_w$ (e.g. at $T = 0.80 T_C$ in figure 3(a)
and $T = 0.82T_{C}$ in figure 3(b)). This observation 
is due to the operation of finite size effects causing 
the effective location of the prewetting temperature to be 
slightly below \( T_{w} \) (see eq.(\ref{eq:Tfinsize})). 
Eventually, this result may also indicate that the critical 
temperature for damage spreading lies below \( T_{w} \), as will
be discussed below. 
%So, in order to unify the notation, hereafter
%$T_{D}(L)$ will be identify as the size dependent ``critical''
%temperature for damage spreading.  

The survival probability of the damage $P(t)$ is shown in figure 4, 
as a function of time $t$. It is observed that $P(t)$ tends to vanish 
for $T < T_{D}(L)$, while it reaches a stationary value for $T > T_{D}(L)$. 
So, it is possible to define the stationary survival probability $P_{stat}$ 
as the asymptotic limit $P_{stat} \equiv P(t \rightarrow \infty)$, 
(see figure 5(a)). In this case, finite-size effects are also observed. 
In fact, the step function expected for the $P_{stat}(T)$ becomes
rounded for lattices of size $L=12$ and $M=601$.
However, the damage  
%wetting 
%!!! OJO no sera la T de damage!!!!!
temperature $T_{D}(L)$ can be determined as 
the mean value of the rounded step.

Figure 6(a) shows log-log plots of the width of the damage profile
$\langle \omega _{x}(t)\rangle$ {\it vs} $t$ obtained 
taking $h = 0.5$ and different temperatures. 
It is observed that for $T > T_{D}(L)$ ( $T < T_{D}(L)$)
the width exhibits an upward (downward) curvature. However,
a straight line behavior emerges for a certain temperature
that is identified as $T_{D}(L) \approx 0.78$. Based on these
observations, the following scaling law is proposed,  
\begin{equation}
\langle \omega _{x}(t)\rangle \propto t^{\alpha } , 
\end{equation}
\noindent where $\alpha$ is an exponent. The best fit of the data 
gives $\alpha = 0.80 \pm 0.01$.

It is worth mentioning that a different behavior has been observed 
for zero surface fields ($h = 0$) close to the
critical temperature \cite{alba4}, as shown in figure 6(b). In this limit
the critical temperature for damage spreading is well known,
namely $T_{D}(h = 0, L = \infty) \simeq 0.992 T_{C}$ \cite{grass2,grass3}.
However, in this case a crossover between a short-time (ST) regime
and a long-time (LT) regime is found close to $t \simeq 350$ mcs.
For $t < 350$ mcs the data is consistent with $\alpha _{ST}=0.517\pm 0.005$,
while for $t > 350$ mcs one has $\alpha _{LT} = 0.79\pm 0.01$. 
This results also suggests that a diffusive-like behavior,
as expected for $\alpha = 1/2$, dominates the short-time regime,
while the propagation of the perturbation becomes faster 
within the long-time regime. 

The observed difference between the cases $h > 0$ and
$h = 0$ can be understood with the aid of the snapshot 
configurations shown in figures 2(a-c) and 2(d), respectively.
In fact, when the interface is most likely localized
at the center of the strip running parallel to the $M$-direction,
where the surface magnetic fields are applied, one expects
an uniform propagation of the perturbation as observed
in figure (6(a)). However, for $h = 0$ the occurrence of 
magnetic domains of opposite orientation exhibiting 
interfaces running perpendicular to the strip determines the
onset of two different time regimes: on the one hand 
the short-time regime, characterized by the (slow) 
propagation of the damage inside the bulk of magnetic 
domains and, on the other hand, the long-time regime with a 
(faster) propagation of the perturbation effectively
catalyzed by large fluctuations occurring at the interfaces
of the domains.   
 
The amplitude of the damage profile in 
the direction parallel to the interface $\langle h_{y}(t)\rangle$,
as shown in figure 7, also exhibits a power-law behavior of the type:
\begin{equation}
\langle h _{y} (t) \rangle \propto t^{\lambda} ,
\label{hy}
\end{equation}
\noindent where $\lambda$ is an exponent. A least-square fit of the data 
(see figure 7) gives $\lambda= 0.98 \pm 0.01$, for $T \geq T_{D}(L)$. 
Also, the damage tends to become healed for $T < T_{D}(L)$. 
Simulations performed using bigger lattices shows that this exponent 
tends to $\lambda \approx 1$ for the asymptotic limit, $L \rightarrow \infty$, 
e.g. $\lambda(L=24)=1.03 \pm 0.02$ and $\lambda(L=48)=1.04 \pm 0.05$.

Figure 8 shows damage profiles $P_{y}(i,t)$ obtained at different 
times, and close to $T_{D}(L)$. For the absence of surface
fields and using open boundary conditions (figure 8(a)), 
the homogeneous propagation of the damage is observed. However,
considering competing surface fields the damage 
tends to propagate in the region where the interface between magnetic 
domains is located, namely at the central rows of the 
film ($i \sim L/2$), as shown in figure 8(b). 
This fact is in qualitative  agreement 
with the results discussed for $\langle h _{y}(t)\rangle$.
Moreover, it is observed that the effect of the magnetic fields 
at the walls is to slow down the propagation of the damage.
These observations are consistent with the fact that 
large fluctuations capable to catalyze the propagation of the
damage are present along the interface, while at the walls and due
to the applied magnetic fields, the orientation of the 
spins is more stable.   

It should be noted that due to the finite size of the
lattice, the width of the damage profile 
\( \langle \omega _{y} (t) \rangle \), 
remains constant close to \( \frac {L} {2} \), preventing a 
quantitative analysis. 

Based on the study of $D(t)$, $P(t)$ and the damage 
profiles, the location of the $L-$dependent 
damage spreading critical points $T_{D}(L)$, for a fixed 
field $h=0.5$, have been estimated as follows:
$T_{D}(L=12) = 0.79(1) T_{C}$,
$T_{D}(L=24) = 0.81(1) T_{C}$ and $T_{D}(L=48) = 0.83(1) T_{C}$.
Based on these results and assuming that $T_{D}(L)$ obeys
a scaling law such as equation (\ref{eq:Tfinsize}), the 
extrapolation $T_{D}(\infty) = 0.84 \pm 0.01$ has been
obtained, as shown in figure 9. This value is close 
(but smaller) to the wetting critical temperature
$T_w(h = 0.5) = 0.863$, as evaluated using eq.(\ref{eq:abrah})
(see also figure 1). It is known that one method to locate the 
$L-$dependent critical temperatures for the localization-delocalization
transition ($T_{w}(L)$), which is the precursor of the true 
wetting temperature in  
the thermodynamic limit, is the evaluation of the maximum of the 
susceptibility, i.e. the fluctuations of the order 
parameter \cite{eva}. Figure 9 also shows that a plot of
$T_{w}(L)$ {\it vs} $L^{-1}$ (see equation (\ref{eq:Tfinsize}))
extrapolates to $T_{w}(L\rightarrow \infty) = 0.866$, in good
agreement with the exact value. Furthermore, 
one has that $T_{D}(L) < T_{w}(L)$.

At this stage it is concluded that, on the one hand, it is hard to establish 
an unambiguous difference between $T_{D}$ and $T_{w}$, while on the other
hand, it is out of any doubt that the onset of damage propagation 
is related to the localization-delocalization transition that undergoes the
interface between competing domains, so that
$T_{D}$ is effectively shifted to temperatures far below from  $T_{C}$.   

\subsection*{IV.2 The case $T = 0.9 T_C$, and $h$ variable.}

In this case, the obtained results are fully consistent with those
obtained keeping the field constant and varying the temperature.
In fact, the total damage $D(t)$ grows up for $h > h_{D}(L)$ 
with exponent $\eta =0.91 \pm 0.02$ (see figure 10), where
$h_{D}(L)$ is $L-$dependent critical field for damage spreading.  
Also, it is observed that for lattices of size $L = 12$, the damage spreading
transition is markedly rounded by finite-size effects (figure 10(a)), 
while for $L = 24$ that transition is better defined (figure 10(b)).
 
The stationary survival probability of the damage $P_{stat}$  
can also be determined, as shown in figure 5(b) as a function 
of the surface field $h$. The fact that the stepped shape of
$P_{stat}$ is better defined in figure 5(b) than in figure 5(a)
is an intriguing feature that remains to be clarified.

%may be due to the fact that for $T = 0.9$ and varying $h$,
%the phase diagram (see figure 1) is crossed almost perpendicularly.
  
The width of the damage profile also behaves according to a power
law (see equation (\ref{eq:pwlaw}) with an exponent 
$\alpha =0.81 \pm 0.01$ (see figure 11), in full agreement with 
our previous estimation given by $\alpha =0.80 \pm 0.01$.

It is found that the amplitude of the profile, 
$\langle h_{y} (t) \rangle$, also exhibits a power law 
close to $T_{D}$ (figure 12). In fact, the exponent $\lambda$
(see equation (\ref{hy}), has been determined given: 
$\lambda = 1.02 \pm 0.01$, in full agreement with our estimation
obtained keeping constant the surface fields, namely
$\lambda = 0.98 \pm 0.01$.

It has been observed that the amplitude along the parallel 
direction and the width of the profiles across the perpendicular 
direction ($\langle h_{x}(t)\rangle$ and $\langle \omega_{y}(t)\rangle$, 
respectively), exhibit the same behavior than in the 
previous case with constant surface fields, so the
corresponding figures are omitted for the sake of space.

Finally, our estimations of the size dependent 
damage spreading critical fields 
$h_D(L)$ are given by: $h_D(L = 12) = 0.31(2)$,  
$h_D(L = 24) = 0.34(1)$ and $h_D(L = 48) = 0.35(1)$. 

By comparison to equation (\ref{eq:Tfinsize}) the following finite size
scaling anzats can be expected to hold for the wetting critical field  
\begin{equation}
h_{c}(L) -  h_{c}(\infty) \sim constant \times L^{-1} ,
\label{eq:hfinsize}
\end{equation}
\noindent where $h_{c}(\infty)$ is the position of the critical point 
in the thermodynamic limit. Plotting the obtained data for the
damage spreading critical fields according 
to equation (\ref{eq:hfinsize})(figure 13), 
the extrapolation to $L \rightarrow \infty$ gives 
$h_{D}(L = \infty) = 0.37 \pm 0.02$, while the exact value 
for the wetting critical point at $T = 0.9T_{c}$, as obtained using 
equation (\ref{eq:abrah}), is $h_{c} = 0.42906$. 
As in the previous case, the location of the
damage spreading critical point lies slightly inside the
non-wet region of the phase diagram (see figure 1).

\subsection*{IV.2 Complementary studies.}    
	
In order to confirm the findings of the previous sections,
additional simulations were performed at $T=0.75T_{C}$
and $T= 0.95T_{C}$ 
and varying the surface fields. All the obtained results are 
consistent with the previous findings. In summary, applying
the scaling anzats given by Equation (\ref{eq:hfinsize}), 
the critical field for damage spreading 
$h_{D}(\infty) = 0.62 \pm 0.01$ 
($h_{D}(\infty) = 0.15 \pm 0.01$) 
was obtained, as compared with the exact result for the 
wetting critical point (e.g. equation (\ref{eq:abrah})) 
given by $h_{c}(T=0.75 T_{C})=0.6652$ ($h_{c}(T=0.95 T_{C})=0.3053$).
Notice that the former extrapolation is shown in figure 13.

\section*{V. Conclusions}

Based on an extensive numerical study of the damage propagation
close to the wetting transition of the Ising model with competing 
fields it is possible to draw the following main conclusions: 

i) The exponent governing the spreading of the total damage
($D(t) \propto t^{\eta}$, see Eq.(\ref{eq:pwlaw})),
in the presence of interfaces in the direction parallel to the 
propagation of the perturbation, namely $\eta \cong 0.90 \pm 0.02$,
is greater than the obtained in absence of interface, 
namely $\eta \cong 0.471 \pm 0.005$ \cite{montani,alba4}. So, clearly  
the interface causes the enhancement of the propagation 
of the perturbation.

From a more qualitative point of view, it is expected that 
the propagation of damage may be favored by interfaces 
because precisely around these regions of the sample
one has the largest fluctuations in the orientation of the spins. 
Within this scenario the spreading of the damage is expected 
to be faster along the direction parallel 
to the interface where the fluctuations can uniformly be propagated. 
In contrast, the spreading along the direction perpendicular 
to the interface may be slowed down by the periods where 
the damage has to cross over the bulk of the domains \cite{alba4}.

ii) Clear evidence on the shift of the critical damage temperature
is reported. In fact, while it is well known that 
$T_{D}(h=0) \approx 0.992T_{C}$ in the absence of surface 
magnetic fields \cite{grass2,grass3}, our finding show 
that  $T_{D}(h>0) < T_{D}(h = 0)$ when competing surface 
fields are applied. Furthermore, all the damage spreading critical points 
extrapolated to the thermodynamic limit lie systematically
within the non-wet phase of the wetting phase diagram (figure 1).
These results suggest that the wetting and the damage spreading 
critical curves are different and point out that the former
should be slightly shifted towards the non-wet phase. 

We expect that this numerical work may contributed 
to the understanding of 
the role of interfaces in the propagation of damage.
The overall understanding of these complex physical 
phenomena addresses a significant theoretical and experimental
challenge. Such kind of studies are stimulated by many 
practical and technological applications as e.g. 
the design and construction of high quality magnetic materials.

\vskip 1.0 true cm
{\bf  ACKNOWLEDGMENTS}. This work is  supported  financially by
CONICET, UNLP, and ANPCyT (Argentina). MLRP acknowledge the 
National Academy of Exact Sciences (Argentina) for the grant of a fellowship.

\newpage

%{\bf Figure Captions}.

\begin{figure}
%\narrowtext
\centerline{{\epsfysize=5.0in \epsffile{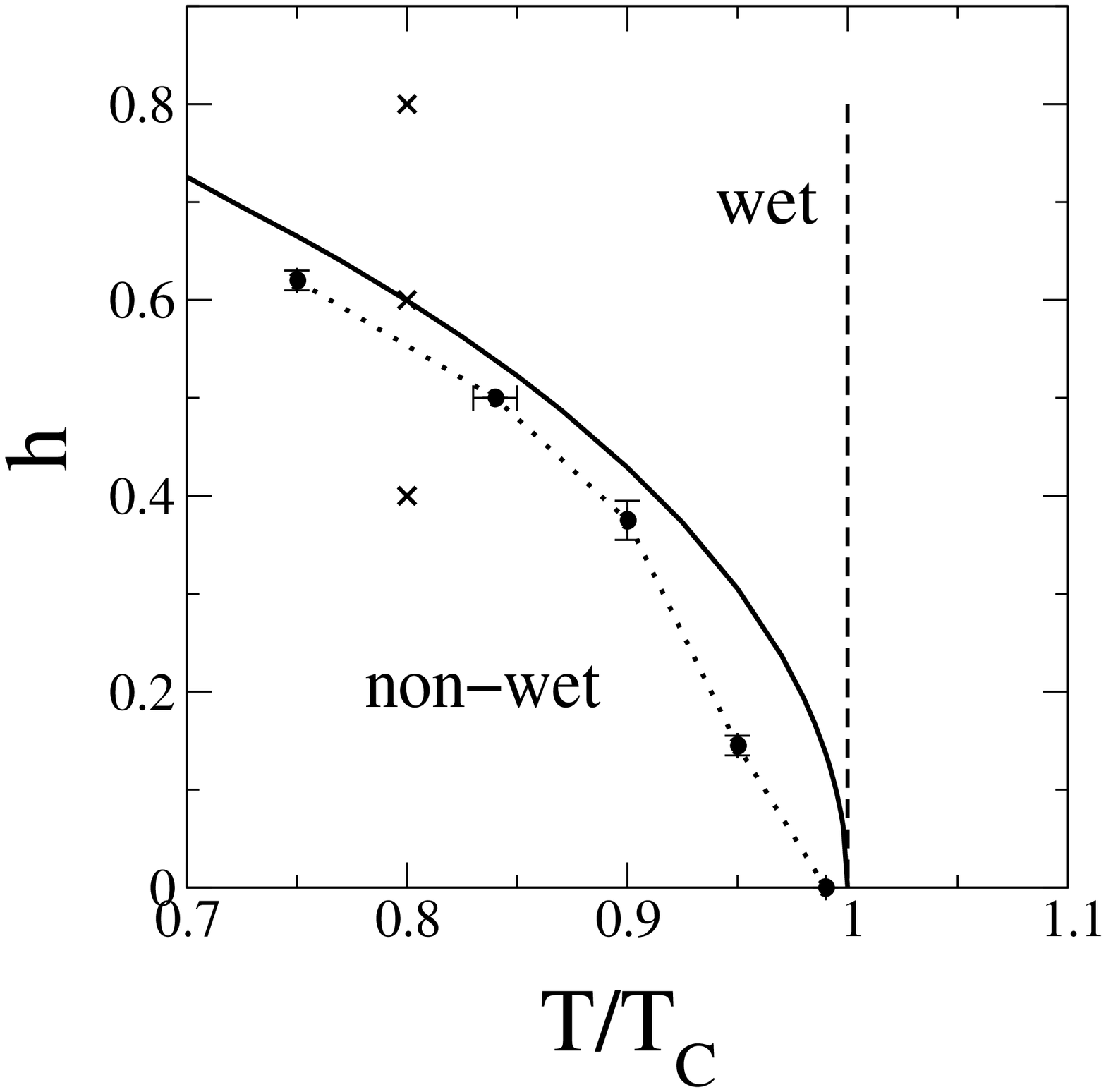}}}
\vskip 1.0 true cm
\caption{Wetting phase diagram ($h$ {\it vs} $T/T_{C}$) 
corresponding to a 
semi-infinite Ising strip with opposite short range surface fields. 
The solid curve is the exact critical line, as obtained by 
Abraham \cite{abra1}. The vertical dashed line is the boundary 
between the disordered phase at right hand side and the ordered 
one at the left hand side. The results obtained for the damage spreading 
transition are shown as full circles, and the dotted line
has been drawn to guide the eyes. Crosses indicate the three points, 
of the phase diagram, where the snapshots pictures shown 
in figure 2 were obtained.}
\label{FIG.1}
\end{figure}

\newpage

\begin{figure}
%\narrowtext
\centerline{{\epsfysize=5.0in \epsffile{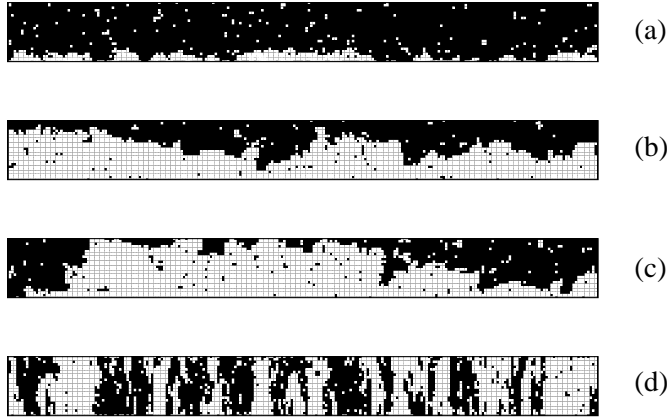}}}
\vskip 1.0 true cm
\caption{Snapshots configurations obtained after $10^4$ mcs 
in confined geometries of size $L=24$ and  $M=1200$.
(a)-(c) Typical spin configurations when short-range fields with opposite 
signs are applied at bottom and top files of the lattice, respectively. 
In these cases, the snapshots are taken at \( T=0.80\, T_{C} \) 
and different surface fields: (a) $h=0.4$, 
within the non-wet phase; (b) $h=0.6$, near the critical wetting curve, 
and (c) $h=0.8$, within the wet phase. The location of these 
points are shown in the phase diagram of figure 1.
(d) Typical equilibirum configuration obtained in absence 
of magnetic fields, applying open boundary conditions, and 
close to $T_C$, $T=0.99T_C$. Note that the horizontal coordinate 
has been reduced by a factor of five in comparison with the 
vertical one for the sake of clarity of the picture.
Sites taken by downs spins are shown in black while up spins 
are left white.}
\label{FIG.2}
\end{figure}

\newpage

\begin{figure}
%\narrowtext
\centerline{{\epsfysize=4.0in \epsffile{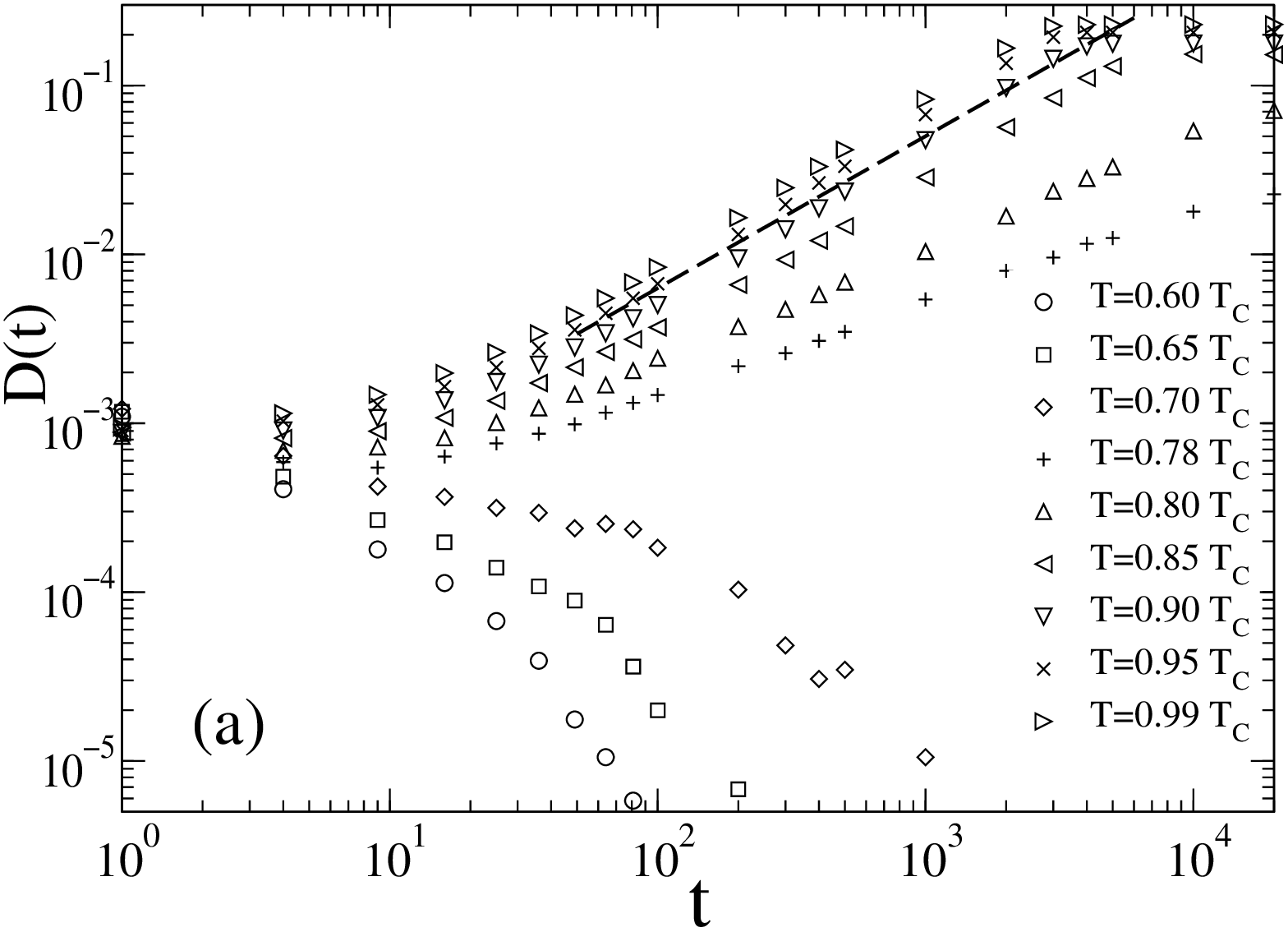}}}
\centerline{{\epsfysize=4.0in \epsffile{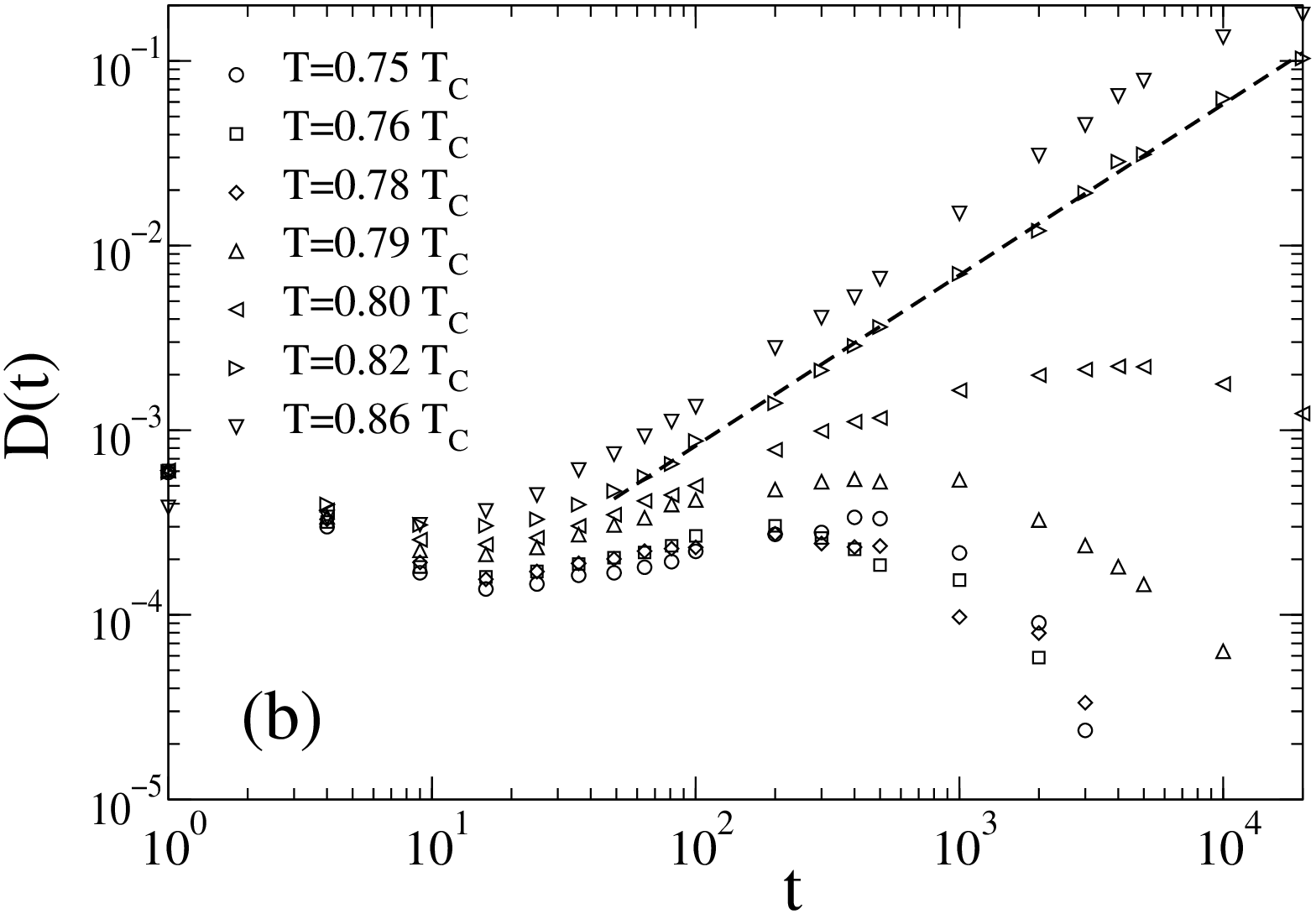}}}
\vskip 1.0 true cm
\caption{Log-log plots of $D(t)$ {\it vs} $t$ obtained 
at different temperatures as indicated in the figure, applying 
short-range fields of magnitude $h = 0.5$, and using lattices of size: 
(a) $L=12, M=601$, (b) $L=24, M=1201$. 
The dashed lines have slopes $\eta = 0.90$ and have been drawn 
for the sake of comparison.}
\label{FIG.3}
\end{figure}

\newpage

\begin{figure}
%\narrowtext
\centerline{{\epsfysize=5.0in \epsffile{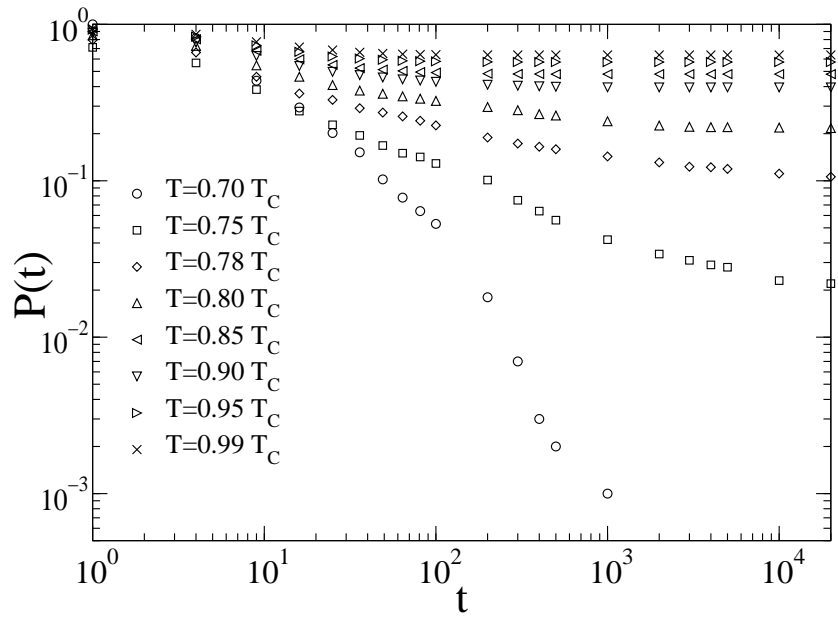}}}
\vskip 1.0 true cm
\caption{Log-log plots of $P(t)$ {\it vs} $t$ obtained
at different temperatures as indicated in the figure, and using 
lattices of size $L=12$ and $M=601$. Short-range fields of magnitude 
$h=0.5$ are applied at the walls of the sample.}
\label{FIG.4}
\end{figure}

\newpage

\begin{figure}
%\narrowtext
\centerline{{\epsfysize=4.0in \epsffile{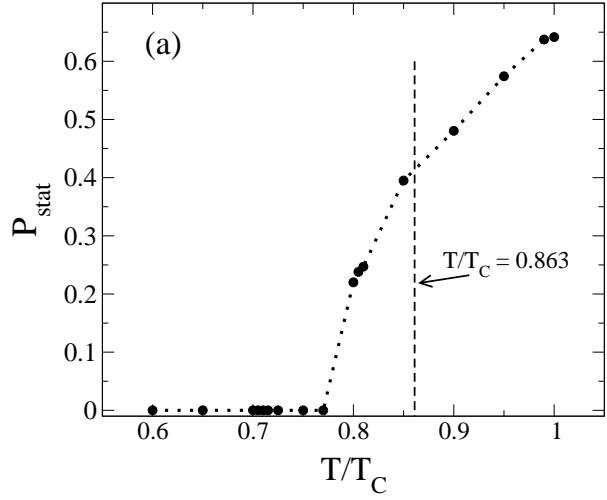}}}
\centerline{{\epsfysize=4.0in \epsffile{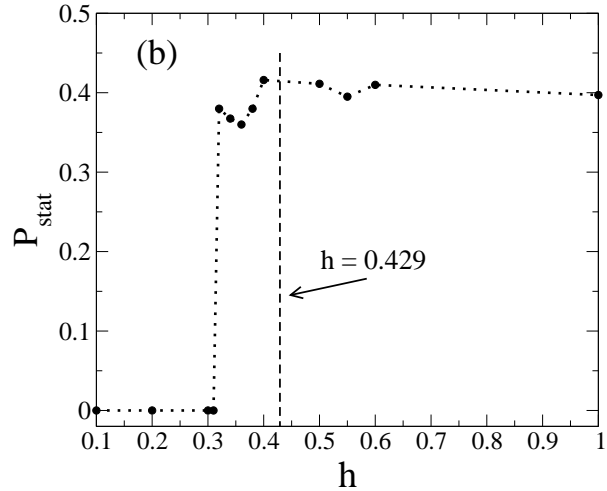}}}
\vskip 1.0 true cm
\caption{Plots of the stationary survival probability 
$P_{stat} \equiv P(\rightarrow \infty)$ as obtained using lattices 
of size $L=12$, $M=601$. 
(a) $P_{stat}$ {\it vs } $T/T_C$ obtained taking short-range 
fields of magnitude $h=0.5$ (the dotted line has been drawn to 
guide the eyes). The vertical dashed line indicates 
the exact critical wetting point: $T_{w}(h_{w}=0.5)=0.863 T_C$.
(b) $P_{stat}$ {\it vs} $h$ obtained taking a fixed temperature 
$T=0.90T_{C}$ (the dotted line has been drawn to guide the eyes).
The vertical dashed line corresponds to the exact wetting critical 
point: $h_{w}(T_{w}=0.90T_{C})  = 0.4291$.}
\label{FIG.5}
\end{figure}

\newpage

\begin{figure}
%\narrowtext
\centerline{{\epsfysize=4.0in \epsffile{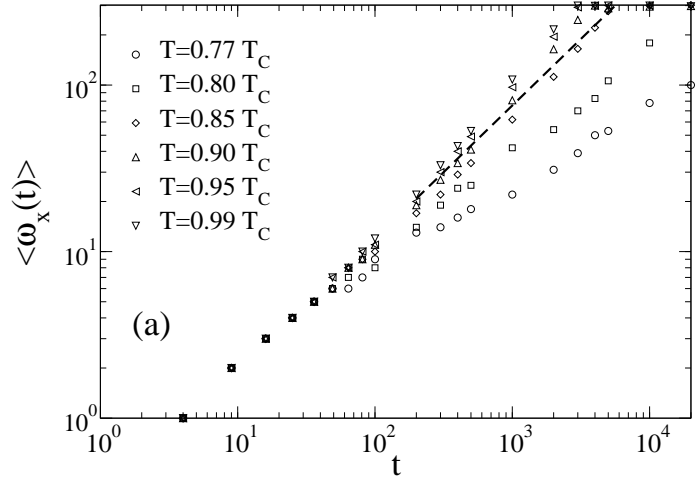}}}
\centerline{{\epsfysize=4.0in \epsffile{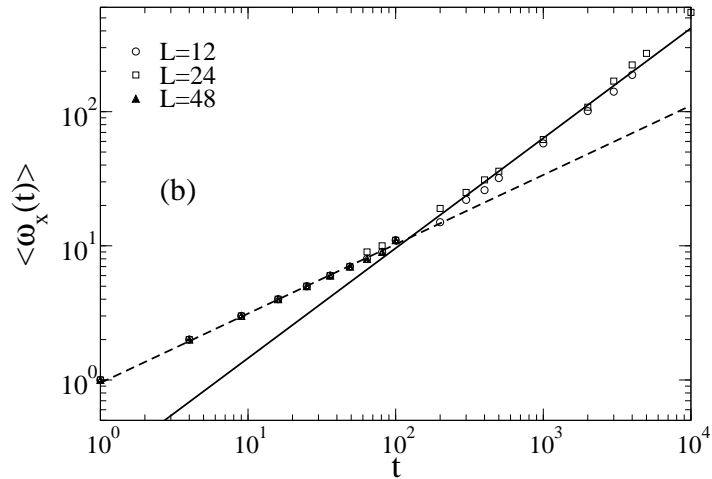}}}
\vskip 1.0 true cm
\caption{Log-log plots of $ <\omega_{x}(t)>$ {\it vs} $t$ 
obtained at different temperatures, as indicated in the figure, 
using lattices of size $L=12$ $M=601$. 
(a) Results obtained taking short-range fields of magnitude $h=0.5$. 
The dashed line has slope $\alpha=0.80$.
(b) Results obtained in the absence of magnetic fields \cite{alba4}
and using lattices of different size, with $M = 50 \times L$,
as indicated in the figure. 
The dashed line has slope $\alpha_{ST}=0.52$, for $t<350$, 
and the full line has slope $\alpha_{LT}=0.78$, 
for $t>350$, respectively. More details in the text.}
\label{FIG.6}
\end{figure}

\newpage

\begin{figure}
%\narrowtext
\centerline{{\epsfysize=5.0in \epsffile{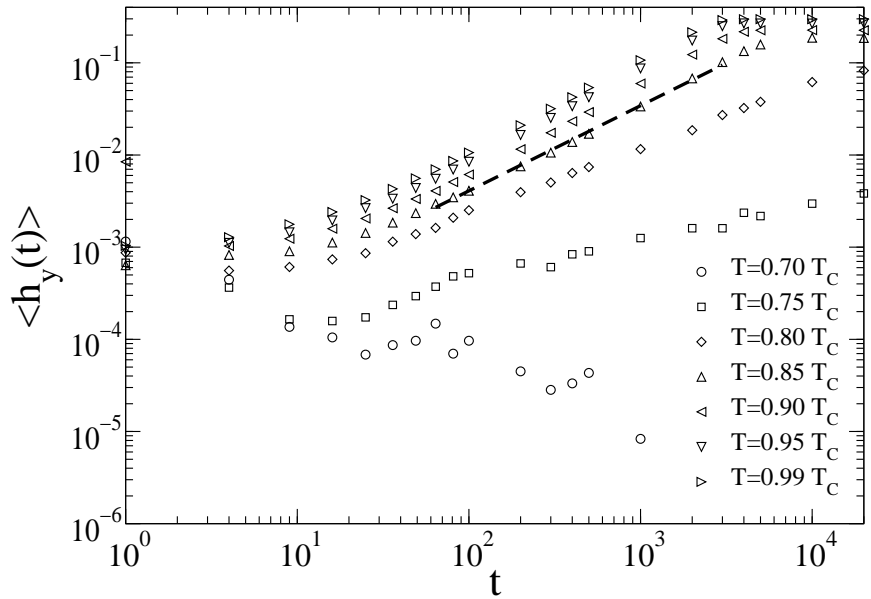}}}
\vskip 1.0 true cm
\caption{Log-log plots of  $<h_{y}(t)>$ {\it vs} $t$ obtained 
at different temperatures, as indicated in the figure, 
using lattices of size $L=12$ $M=601$, and applying short-range 
fields of magnitude $h = 0.5$. The dashed line has slope $\lambda=0.98$.}
\label{FIG.7}
\end{figure}

\newpage

\begin{figure}
%\narrowtext
\centerline{{\epsfysize=5.0in \epsffile{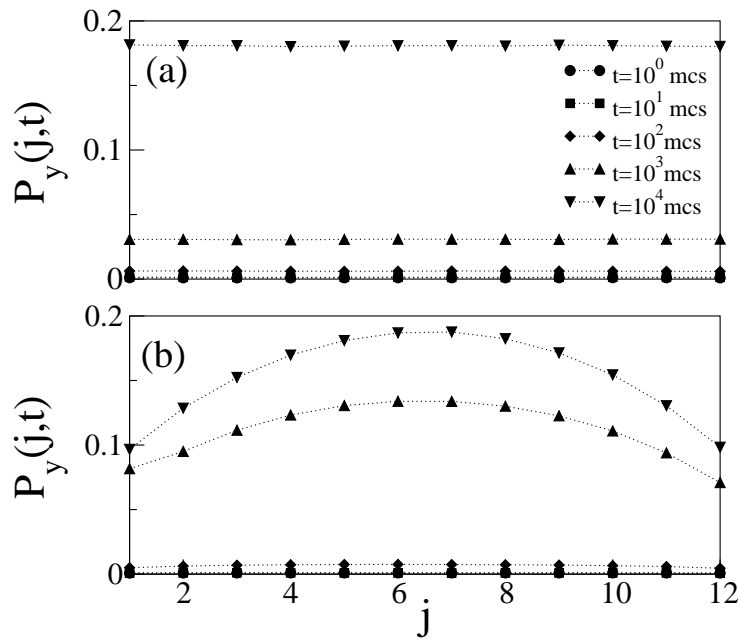}}}
\vskip 1.0 true cm
\caption{Plots of the damage profile measured along 
the $L-$direction and obtained at different times as indicated
in the figure. Results are averaged over \( 10^{3} \) samples, 
using lattices of size $L=12$, $M=601$.  
(a) $T=0.98T_{C}$ and open boundary conditions, 
(b) $T=0.861T_{C}$ and short-range 
fields $\left| h_{1}\right| =\left| h_{L}\right| =0.5$.}
\label{FIG.8}
\end{figure}

\newpage

\begin{figure}
%\narrowtext
\centerline{{\epsfysize=5.0in \epsffile{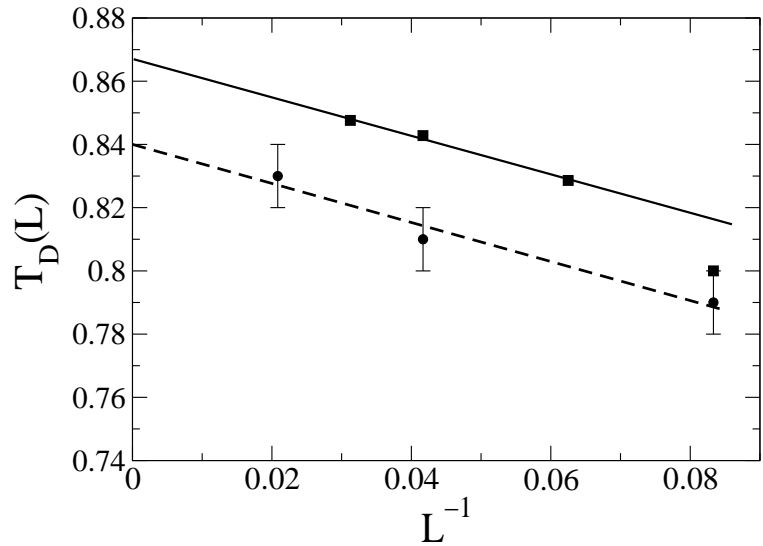}}}
\vskip 1.0 true cm
\caption{Damage spreading critical points $T_{D}(L)$ 
as a function of $L^{-1}$, obtained for lattices of different size 
($L=12, 24, 48$, with $M = 50 \times L$), and applying short range
surfaces fields of magnitude $h=0.5$ (full circles). The dashed 
line extrapolates to the critical temperature $T_{D}/T_{C}=0.84\pm 0.01$. 
Full squares correspond to the interface localization-delocalization 
transitions $T_{w}(L)$ as obtained for $h=0.5$ and using the criteria of 
the maximum of the susceptibility \cite{eva}. The full line
extrapolates to the wetting critical temperature  $T_{w}/T_{C}=0.866$.}
\label{FIG.9}
\end{figure}

\newpage

\begin{figure}
%\narrowtext
\centerline{{\epsfysize=4.0in \epsffile{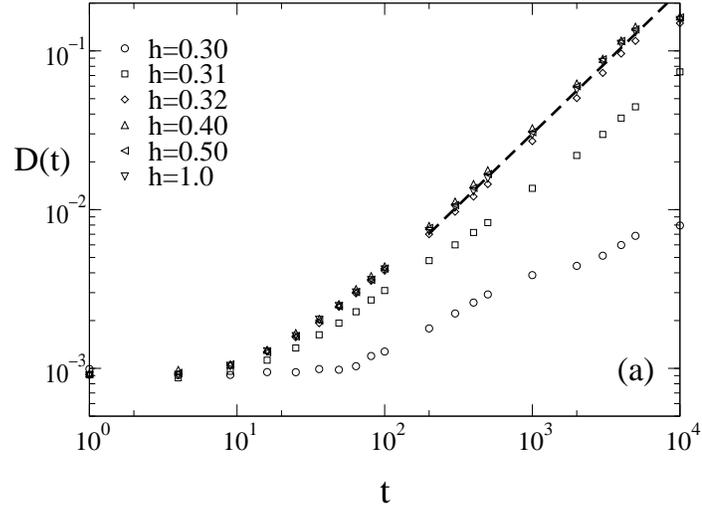}}}
\centerline{{\epsfysize=4.0in \epsffile{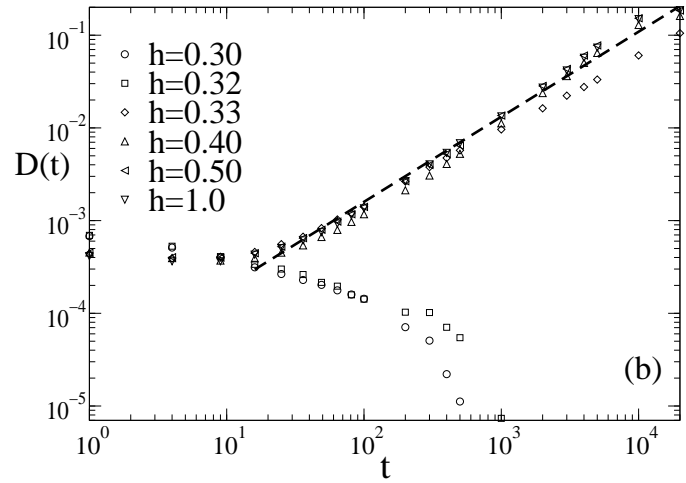}}}
\vskip 1.0 true cm
\caption{Log-log plots of $D(t)$ {\it vs} $t$ obtained 
using short-range magnetic fields of different magnitude as indicated 
in the figure, at fixed temperature $T=0.90 T_C$, and using 
lattices of size: (a) $L=12$, $M=601$, (b) $L=24$, $M=1201$. 
The dashed lines have slopes $\eta \sim 0.90$.}
\label{FIG.10}
\end{figure}

\newpage

\begin{figure}
%\narrowtext
\centerline{{\epsfysize=5.0in \epsffile{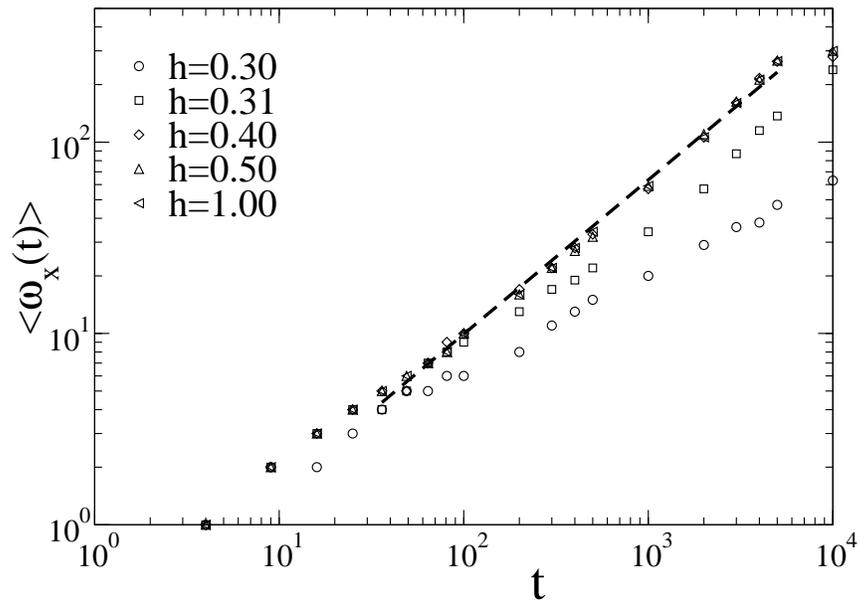}}}
\vskip 1.0 true cm
\caption{Log-log plots of $< \omega_{x}(t)>$ {\it vs} $t$ 
obtained for short-range magnetic fields of different magnitude, 
as indicated in the figure, using lattices of size $L=12$ $M=601$, 
and at fixed temperature $T = 0.90 T_C$. The dashed line has 
slope $\alpha=0.81$.}
\label{FIG.11}
\end{figure}

\newpage

\begin{figure}
%\narrowtext
\centerline{{\epsfysize=5.0in \epsffile{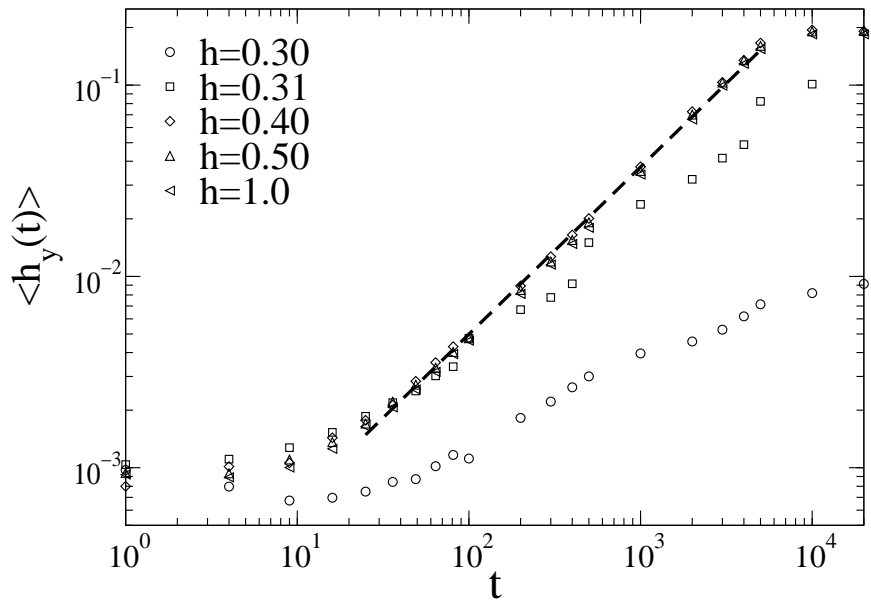}}}
\vskip 1.0 true cm
\caption{Log-log plots of $<h_{y}(t)>$ {\it vs} $t$ obtained 
at different short-range magnetic fields, as indicated in the figure, 
using lattices of size $L=12$ $M=601$, and at fixed temperature $T=0.90 T_C$.
The dashed line has slope $\lambda=1.02$.}
\label{FIG.12}
\end{figure} 

\newpage

\begin{figure}
%\narrowtext
\centerline{{\epsfysize=5.0in \epsffile{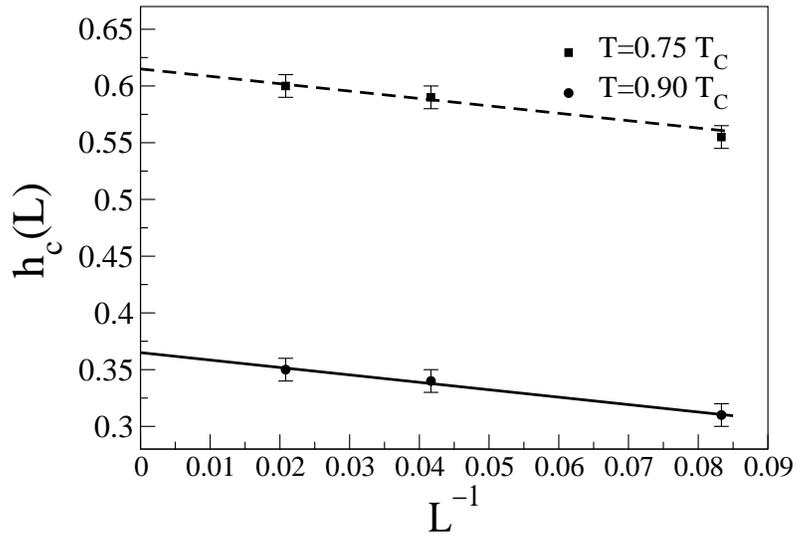}}}
\vskip 1.0 true cm
\caption{Damage spreading critical fields $h_{D}(L)$ as a 
function of $L^{-1}$, obtained for lattice of different size 
($L=12, 24, 48$), and at different temperatures, as indicated in 
the figure. The dashed (full) line extrapolates to the critical field
$h_{D}(L \rightarrow \infty, T=0.75 T_{C})=0.62\pm 0.01$ 
($h_{D}(L \rightarrow \infty, T=0.90 T_{C})=0.37\pm 0.02$). 
Both lines have the same slope.}
\label{FIG.121}
\end{figure}

\end{document}